\documentclass[12pt]{article}

\usepackage{amsmath,amssymb,theorem,latexsym,graphics}

\usepackage[all]{xy}
\usepackage{color}
\newtheorem{theo}{Theorem}[section]

{\theorembodyfont{\rmfamily} }
{\theorembodyfont{\rmfamily} \newtheorem{ex}[theo]{Example}}
{\theorembodyfont{\rmfamily} }
{\theorembodyfont{\rmfamily} }

\DeclareFontFamily{U}{rsf}{}
\DeclareFontShape{U}{rsf}{m}{n}{
  <5> <6> rsfs5 <7> <8> <9> rsfs7 <10->  rsfs10}{}
\DeclareMathAlphabet{\mathscr}{U}{rsf}{m}{n}
\newcommand{\mycal}[1]{\mathscr{#1}}
\newcommand{\op}[1]{\operatorname{#1}}

\newcommand{\oX}{{\overline X}}
\newcommand{\wX}{{\widetilde{X}}}

\newcommand{\CF}{{\cal{F}}}

\newcommand{\half}{\frac{1}{2}}

\newcommand{\bM}{{\boldsymbol{M}}}
\newcommand{\bL}{{\boldsymbol{L}}}

\newcommand{\ra}{{\longrightarrow}}

\newcommand{\be}{\begin{equation}}
\newcommand{\ee}{\end{equation}}
\newcommand{\IP}{{\mathbb P}}
\newcommand{\IC}{{\mathbb C}}
\newcommand{\IZ}{{\mathbb Z}}

\newcommand{\CL}{{\mathcal L}}
\newcommand{\CU}{{\mathcal U}}

\newcommand{\CX}{{\mathcal X}}
\newcommand{\IF}{{\mathbb F}}
\newcommand{\oCX}{{\overline{\mathcal X}}}
\newcommand{\oDelta}{{\overline \Delta}}
\newcommand{\os}{{\overline s}}

\newcommand{\hX}{{\widehat X}}

\newcommand{\wm}{{\widetilde m}}
\newcommand{\wbM}{{\widetilde {\boldsymbol{M}}}}

\newcommand{\wGamma}{{\widetilde \Gamma}}

\newcommand{\bD}{{\boldsymbol D}}
\newcommand{\wCF}{{\widetilde{\mathcal F}}}
\newcommand{\ret}{\boldsymbol{\op{ret}}}
\begin{document}

\title{Geometric transitions and mixed Hodge structures}
\author{D.-E. Diaconescu$^{\flat}$, R. Donagi$^{\dagger}$, and T. Pantev$^{\dagger}$\\
$^\flat$ {\small New High Energy Theory Center, Rutgers University}\\
{\small 126 Frelinghuysen Road, Piscataway, NJ 08854}\\
$^{\dagger}$ {\small Department of Mathematics, University of Pennsylvania} \\
{\small David Rittenhouse Lab., 209 South 33rd Street,
Philadelphia, PA  19104-6395}\\
}
\date{June 2005}

\maketitle

\begin{abstract}
We formulate and prove a {\bf B}-model disc level large $N$ duality
result for general conifold transitions between compact Calabi-Yau
spaces using degenerations of Hodge structures.
\end{abstract}

%\tableofcontents

\section{Introduction}

Large $N$ duality is a relation between open and closed string
theories on two different Calabi-Yau manifolds connected by an
extremal transition\cite{GV:corresp}.  This relation was originally
formulated in the context of topological {\bf A}-model for a local
conifold transition \cite{GV:corresp} and it was extended to
more general noncompact toric Calabi-Yau manifolds in
\cite{AV:gtwo,DFG:open,DFG:delPezzo,AMV:all}.

This paper is part of a long term project aimed at understanding large
$N$ duality for extremal transitions between compact Calabi-Yau
manifolds. A first step in this direction has been made in
\cite{DF:largeN} for topological {\bf A}-models. Here we will be
concerned with large $N$ duality in the topological {\bf B}-model.
Open-closed duality for topological {\bf B}-strings was first
developed in \cite{DV:matrix,DV:geometry} for a special class of
noncompact toric Calabi-Yau manifolds employing a remarkable relation
between holomorphic Chern-Simons theory and random matrix models.

In contrast with the {\bf A}-model, the topological {\bf B}-model on
compact Calabi-Yau spaces has not been given so far a rigorous
mathematical description. However it is well known that the genus zero
topological closed string amplitudes can be expressed in Hodge
theoretic terms using the formalism of special geometry. On the other
hand disc level topological open string amplitudes associated to
D-branes wrapping curves in Calabi-Yau threefolds can also be given a
geometric interpretation in terms of Abel-Jacobi maps
\cite{DT:gauge,EW:branes,HC:obs,AV:discs}.  Higher genus amplitudes do
not have a pure geometric interpretation. In principle one would have
to quantize Kodaira-Spencer theory coupled to holomorphic Chern-Simons
theory on a compact Calabi-Yau space, which is a very hard task at
best.

In this paper we will formulate and prove a first order {\bf B}-model
duality statement for general conifold transitions between compact
Calabi-Yau spaces.  By first order duality we mean a correspondence
between topological disc amplitudes on the open string side and first
order terms in a suitable expansion of the holomorphic prepotential on
the closed string side. The expansion is taken around an appropriate
stratum parameterizing nodal Calabi-Yau spaces that admit a projective
crepant resolution.

Using special geometry, in section two we show that the first order
terms in this expansion admit a intrinsic geometric interpretation in
terms of degenerations of Hodge structures.  In section three we will
show that the first order duality statement follows from a Hodge
theoretic result relating two different mixed Hodge structures. The
main element in the proof is the Clemens-Schmid exact sequence.

A connection between mixed Hodge structures and {\bf B}-model
topological disc amplitudes on toric Calabi-Yau manifolds has been
previously developed in \cite{LMW:hol,LMW:mixed,BF:PF}. This machinery
has been applied to first order large $N$ duality for toric Calabi-Yau
manifolds in \cite{BF:Bmodel}. Our approach is different and can be
used to extend the ${\bf B}$-model large $N$ duality beyond
disc level. Some progress along these lines for an interesting class
of noncompact transitions is reported in the companion paper
\cite{DDDHP:largeN}.

\

\medskip

\noindent {\it Acknowledgments.} We are very grateful to Bogdan
Florea and Antonella Grassi for collaboration at an early stage of
the project and many useful discussions. The work of D.-E. D. has
been partially supported by an Alfred P. Sloan fellowship. R.D. is
partially supported by NSF grant DMS 0104354 and FRG grant 0139799
for ``The Geometry of Superstrings''. T.P. is partially supported
by NSF grants  FRG 0139799 and DMS 0403884.

\section{ {\bf B}-Model transitions and periods} \label{s-periods}

In this section we discuss the first order behavior of {\bf B}-model
geometric transitions associated to conifold singularities of
Calabi-Yau spaces. For us a Calabi-Yau space will be a (possibly
singular) complex Gorenstein quasi-projective variety $X$ which has a
trivial canonical class. In addition, in the singular case we will
require that $X$ has a K\"{a}hler crepant resolution.

This convention is somewhat broader than the usual notion of a
Calabi-Yau space used in physics, where one requires that $X$ is a
complex analytic space equipped with a Ricci flat K\"{a}hler metric.
We will be primarily interested in moduli spaces parameterizing
Calabi-Yau structures. These moduli spaces can have different
components corresponding to different sets of values of the
topological invariants of $X$. A geometric (or extremal) transition is
a process connecting two connected components of the moduli space
through a degeneration. Schematically an extremal transition  is
captured in a diagram
\[
\xymatrix{
& {\wX}_{\wm} \ar[d]  \\
X_{l} \ar@{~>}[r] & {X}_{m}.
}
\]
where $X_{l}$ is a smooth Calabi-Yau, $\xymatrix@1{{X}_{l} \ar
  @{~>}[r] & {X}_{m}}$ is a degeneration of $X_{l}$ to a Calabi-Yau variety
$X_{m}$ having only ordinary double points, and $\wX_{\wm} \to X_{m}$ is
a crepant quasi-projective resolution of $X_{m}$.

\

\bigskip

\noindent
The standard example of such a situation is the local conifold
transition:

\begin{ex} Take $\{ X_{\mu} \}_{\mu \in {\mathbb C}}$ to be the one
    parameter  family of $3$-dimensional affine quadrics
$X_{\mu} = \{ (x,y,z,w) \in {\mathbb C}^{4} | xy - zw = \mu \}$. When
$\mu \to 0$ we get a degeneration $\xymatrix@1{{X}_{\mu} \ar
  @{~>}[r] & {X}_{0}}$ of $X_{\mu}$ to the $3$-dimensional quadratic
cone $xy = zw$. To complete this degeneration to a transition
\[
\xymatrix{
& {\wX}_{0} \ar[d]  \\
X_{\mu} \ar@{~>}[r] & {X}_{0},
}
\]
we take $\wX_{0}$ to be one of the small resolutions of $X_{0}$,
i.e. the blow-up of $X_{0}$ along the Weil divisor in $X_{0}$
corresponding to one of the two rulings in the base ${\mathbb
  P}^{1}\times {\mathbb P}^{1}$ of the cone. For future reference we
note also that as an abstract variety
$\wX_{0}$ is isomorphic to the total space of ${\mathcal O}(-1)\oplus
{\mathcal O}(-1) \to {\mathbb P}^{1}$, and the map $\wX_{0} \to X_{0}$
is the natural map contracting the zero section of ${\mathcal O}(-1)\oplus
{\mathcal O}(-1)$.
\end{ex}

\

\noindent
The geometric transition from $X_{l}$ to $\wX_{\wm}$ {\em changes the
topology}. Indeed, the process of degenerating $X_{l}$ to $X_{m}$
collapses some $3$-spheres in $X_{l}$ to the singular points of
$X_{m}$. These spheres are the vanishing cycles of the degeneration
$\xymatrix@1{X_{l} \ar@{~>}[r] & X_{m}}$. On the other hand, the small
resolution $\wX_{\wm} \to X_{m}$ replaces each singular point of
$X_{m}$ by a copy of ${\mathbb P}^{1} \cong S^{2}$. So in the passage
from $X_{l}$ to $\wX_{m}$ we deleted some $3$ spheres from $X_{l}$ and
glued $2$-spheres in their place. It is important to note that this
process not only changes the topology but also alters the type of
geometry of the Calabi-Yau spaces in question. More precisely the
transition interchanges holomorphic and symplectic data: the
exceptional ${\mathbb P}^{1}$'s in $\wX_{\wm}$ are holomorphic curves,
and conjecturally the vanishing $3$-spheres in $X_{l}$ can be chosen
to be special Lagrangian submanifolds for the K\"{a}hler form on
$X_{l}$.  In particular, in an {\bf A}-model transition one expects a
relationship between open Gromov-Witten invariants on $X_{l}$ (with
boundaries on the vanishing $3$-spheres) and closed Gromov-Witten
invariants on $\wX_{m}$. The precise form of such a relationship is
the content of the {\bf A}-model large $N$ duality which has been
extensively analyzed in the literature, see
\cite{DFG:delPezzo,DFG:open,DF:largeN} and references therein. Here we
propose a geometric description for the {\bf B}-model counterpart of
the large $N$ duality and investigate some of the mathematical and
physical consequences of our proposal.

Suppose $\bL$ is a fixed component of the moduli space of Calabi-Yau
threefolds with at most ODP singularities.  Given a point $l \in \bL$
we write $X_{l}$ for the corresponding Calabi-Yau. We will always
assume that for a general $l \in \bL$, the variety $X_{l}$ is a smooth
(compact or non-compact) Calabi-Yau threefold. In examples we will
often take $X_{l}$ to be complete intersections in some toric variety
since we want to keep track of the family of ${\bf A}$-models mirror
to the topological ${\bf B}$-models specified by the $X_{l}$'s.

More precisely, we will look at the subvariety of $\bL$ parameterizing
singular threefolds with ordinary double points which admit a crepant
projective resolution.  Let $\bM$ be a component of this subvariety
and let $v$ denote the number of ODPs of $X_{m}$ for a general $m \in
\bM$. In particular on a nearby smooth $X_{l}$ we have a collection of
$v$ embedded Lagrangian 3-spheres $L_{1},\ldots, L_{v}$ whose homology
classes $[L_{1}], \ldots, [L_{v}] \in H_3(X_{m}, {\mathbb Z})$ vanish
under a deformation $\xymatrix@1{X_{l} \ar@{~>}[r] & X_{m}}$.  Recall
that \cite[Theorem 2.9]{STY:conifold} in
order for $X_m$ to admit a projective small resolution, we must have at
least one good relation among $[L_{1}],\ldots, [L{_v}]$. That is, in
$H_{3}(X_{l},{\mathbb Z})$ we must have a relation of the form
\[
\sum_{i = 1}^{v} c_{i} [L_{i}] = 0 \in H_{3}(X_{l},{\mathbb Z}),
\text{ with } c_{i} \neq 0 \text{ for all } i = 1, \ldots, v.
\]
Assuming that this is the case, let $r\geq 1$ denote the number of
relations on the vanishing cycles.  Then for a fixed point $m\in M$,
$X_m$ may have finitely many different projective small resolutions
related by flops. This means that the moduli space $\wbM$ of the
resolution is a finite to one cover of the component $\bM$. In the
following we will denote by $\rho : \wbM\to \bM$ the covering map and
by $\wm$ a point of $\wbM$ which projects to $m\in \bM$.  Since above
we have restricted our considerations to the moduli space of
Calabi-Yau threefolds with at most isolated ODP singularities, this
cover is unramified. The branching points of the cover would
correspond to singular threefolds with more complicated singularities
which have been excluded by our definition of the moduli space $\bL$.

The exceptional locus of the resolution $\wX_{\wm} \to X_{m}$ consists
of $v$ smooth $(-1,-1)$ rational curves $C_1,\ldots, C_v$ satisfying
$v-r$ relations in $H_{2}(\wX_{\wm},{\mathbb Z})$. Moreover, we have
(see e.g. \cite{C:dsolids}) the following relations among Betti
numbers of $X_{m}$ and $\wX_{\wm}$:
\[
\begin{split}
b_2(\wX_{\wm}) & = b_2(X_{m})+ r = b_{2}(X_{l}) + r, \\
b_3(\wX_{\wm}) & = b_3(X_{m})-(v-r) = b_{3}(X_{l}) - 2(v-r), \\
b_4(\wX_{\wm}) & = b_4(X_{m}) = b_{4}(X_{l}) + r.
\end{split}
\]
Large $N$ duality conjectures a correspondence between topological
string theories defined on the Calabi-Yau manifolds $X_l$ and
$\wX_\wm$ related by a geometric transition. In this paper we will
study {\bf B}-model transitions, in which case the conjecture predicts
an equivalence between closed topological strings on $X_l$ and
open-closed topological strings on $\wX_\wm$.  In physical terms, the
open closed topological string theory on the small resolution $\wX_\wm$
is defined by wrapping $N_i$ {\bf B}-branes on the exceptional curves
$C_i$.  It is by now well established that in a rigorous framework
{\bf B}-branes should be described by derived objects. However, for
the purposes of the present paper, it suffices to think of a {\bf
B}-brane as an algebraic cycle on $\wX_\wm$ of the form
$\sum_{i=1}^vN_iC_i$.  Furthermore we will restrict our considerations
to homologically trivial D-brane configurations i.e.
\be\label{eq:gauss}
\sum_{i=1}^v N_i [C_i] =0
\ee
where $[C_i]\in
H_2(\wX_\wm,\mathbb{Z})$ denotes the homology class of $C_i$.  In
principle, the open-closed topological ${\bf B}$ model should be well
defined from a physical point of view for any values of the
multiplicities $N_i$.  However it will become clear later that the
above homology condition is required by large $N$ duality.  While
there is no a priori explanation for this condition in topological
string theory, in physical superstring theory, this is a direct
consequence of the Gauss law for Ramond-Ramond flux.  It is quite
interesting that the topological version of large $N$ duality still
requires us to impose the physical Gauss law.

\subsection{Closed Strings} \label{ss:closed}

The central object of study of any topological string theory is the
partition function, which is a generating functional for topological
string amplitudes.  The partition function of the closed topological
${\bf B}$-model on $X_l$ can be written as
\[
\CF_{X_l}^{cl}= \sum_{g=0}^\infty g_s^{2g-2} \CF^g_{X_l}.
\]
The genus $g$ free energy $\CF_{X_l}^g$ is heuristically defined in
terms of functional integrals over moduli spaces of maps from compact
genus $g$ Riemann surfaces to $X_l$. In the {\bf B}-model the
functional integral receives contributions only from degenerate maps,
which sit on the boundary of the moduli space. For genus
zero, the degenerate maps in question are constant maps, and the
functional integral reduces to an ordinary integral on $X_l$
\cite{EW:mirror}. Moreover the genus zero free energy depends
holomorphically on the complex structure parameters of $X_l$.  For
higher genus, the degenerate maps have a more complicated structure
and there is no rigorous mathematical formulation of topological
amplitudes.  In the following we will restrict ourselves to genus zero
topological strings.

Next we will explain the construction of the genus zero free energy
$\CF^0_{X_l}$ and its relation to the special geometry of
the moduli space $\bL$. Since in the {\bf B}-model all physical
correlators depend on the choice of a global holomorphic three-form,
we have to introduce the enlarged moduli space ${\bL}'$ parameterizing
pairs $(X_l,\Omega_l)$ where $\Omega_l$ is a nonzero global
holomorphic three-form on $X_l$. Note that there is a complex
holomorphic line bundle $\CL\to {\bf L}$ so that the fiber $\CL_l$ is
the space of global holomorphic three-forms on $X_l$ for any point
$l\in {\bf L}$.  The enlarged moduli space ${\bL}'$ is isomorphic to
the complement of the zero section in the total space of $\CL$, hence
it has the structure of a holomorphic principal $\IC^\times$-bundle
$\pi: {\bL}' \to \bL$. Let us denote by $\bL_0$ the open subspace of
${\bL}$ parameterizing smooth varieties $X_l$, and by ${\bL}'_0$ its
inverse image in ${\bL}'$. We also write $\bM'$ for the inverse image
of $\bM$ in $\bL'$, and $\wbM'$ for the enlarged moduli space of the
resolution. Note that there is a finite to one unramified cover
$\rho':\wbM'\to \bM'$.

\

\noindent
{\bf Caution:} The previous discussion is somewhat loose. For
instance, the moduli $\bL'$ of pairs $(X_{l},\Omega_{l})$ is the total
space of a line bundle $\mathcal{L} \to \bL$ only if we view $\bL$ as
a stack. More importantly we need to make sure that $\mathcal{L} \to
\bL$ is a line bundle on $\bL - \bL_{0}$ as well. If we have a
universal family $f : \mycal{X} \to \bL$, then $\mathcal{L}$ is the
pushforward $f_{*}\omega_{\mycal{X}/\bL}$ of the relative dualizing
sheaf,  which is locally free by
cohomology and base change. Indeed for this we only need to note that
for a nodal Calabi-Yau $X_{m}$ we have $h^{0}(X_{m}, K_{X_{m}}) =
h^{0}(X_{m},\mathcal{O}) = 1$
and $h^{1}(X_{m},K_{X_{m}}) = h^{1}(X_{m},\mathcal{O}) = 0$.

\

\bigskip

Ideally one would like to define $\CF^0_{X_l}$ as an intrinsic global
geometric object on the moduli space $\bL'$ which can be locally
described as a holomorphic function (for example a section in a
certain line bundle.)  Unfortunately, there is no such intrinsic
construction for $\CF^0_{X_l}$. One can construct the three point
function, or Yukawa coupling, as a global cubic form on $\bL'_0$
\cite{BG:periods,DM:guide}.  The genus zero free energy can only be
defined locally as a primitive of the Yukawa coupling. This
description is of course ambiguous since the Yukawa coupling
specifies only the third derivatives of the free energy. Therefore in
order to obtain a well defined local function we have to make some
choices. Using special geometry (see e.g. \cite{BG:periods,
DF:special}) one can show that a local primitive for the Yukawa
coupling is determined by a choice of splitting of the third homology
$H_3(X_l,\IZ)$ into a direct sum $A\oplus B$ of maximal Lagrangian
sublattices.

Recall that we have chosen $\bL$ to be a component of a moduli space of Calabi-Yau threefolds
with at most isolated ODP singularities. $\bM$ is a subvariety of the discriminant parameterizing
threefolds with a fixed number $v$ of ODPs which admit a crepant projective resolution.
In generic situations, the $(v-r)$ codimensional subvariety $\bM$ of the discriminant can be locally represented as
the intersection locus of $v$ (local) branches of the discriminant. Therefore we can choose an open subset $\CU\subset \bL$
so that $\CU\cap \bM$ is the intersection locus of a collection of Weil divisors
$\bD_1,\ldots, \bD_v$ in $\CU$ so that $v-r$ of them intersect transversely along $\bM$.
Moreover $X_l$ with $l\in \CU$ is singular if and only if $l$ is a point on
\[
\bD = \bD_1 +\ldots + \bD_v.
\]

Let $\CU'$ denote the inverse image of $\CU$ in $\bL'$. In order to write down a local expression for
the genus zero free energy we have to introduce special coordinates $z^\alpha$ on $\CU'$
by choosing a symplectic basis of three-cycles $\{\gamma_{\alpha,l},\gamma^\alpha_l\}$,
$\alpha =1,\ldots h^{1,2}(X_l)+1$ on each
threefold $X_l$, with $l\in \CU \setminus \bD$. The symplectic basis of cycles determines
a splitting $H_3(X_l, \IZ) = A_l\oplus B_l$ where $A_l,B_l$ are complementary maximal Lagrangian sublattices
spanned by the cycles $\{\gamma_{\alpha,l}\}$ and $\{\gamma^\alpha_l\}$ respectively.
Note that we have a monodromy transformation
\be\label{eq:monodromyA}
T_i : H_3(X_l, \IZ) \to H_3(X_l, \IZ), \qquad T_i(\Gamma) = \Gamma + \langle \Gamma, \xi_i\rangle \xi_i
\ee
associated to each component $\bD_i$, $i=1,\ldots, v$ of the discriminant in $\CU$,
where $\xi_i\in H_3(X_l, \IZ)$ is the corresponding vanishing cycle. We have denoted by
$\langle\ ,\ \rangle$ the intersection pairing in $H_3(X_l, \IZ)$.
In order to obtain single valued coordinates on $\CU'$, we have to choose the lattices of
$A$-cycles so that $A_l$ is contained in the fixed locus of $T_i$ for each $i=1,\ldots, v$.
Then the special projective coordinates are given by
\[
%%\label{eq:specialA}
z^\alpha = \int_{\gamma_{\alpha,l}} \Omega_{X_l}
\]
where $\Omega_{X_l}$ is a global holomorphic three-form on $X_l$. Since the cycles
$\gamma_{\alpha,l}$ are fixed by the monodromy transformations
\eqref{eq:monodromyA}, the coordinates $z^\alpha$ extend as single
valued holomorphic functions over the entire open set $\CU'$.

The genus zero free energy -- or, adopting special geometry terminology, the holomorphic prepotential --
is a multivalued holomorphic function on $\CU'\cap \bL_0'$ given by
\be\label{eq:specialB}
\CF^0 = \half \sum_{\alpha=1}^{h^{1,2}(X_l)+1} z^\alpha \Pi_\alpha
\ee
where
\[
\Pi_\alpha = \int_{\gamma^\alpha_l} \Omega_{X_l}
\]
are the periods of the holomorphic three-form on the $B$-cycles $\gamma^\alpha_l$.
We also have the special geometry relations
\be\label{eq:specialC}
\Pi_\alpha = \frac{{\partial \CF^0}}{\partial z^\alpha}.
\ee
which will be useful later in the paper.

\subsection{Open Strings}

Let us now discuss the topological open-closed {\bf B}-model on the
small resolution $\wX_\wm$. In open-closed topological string theory,
one would like to integrate over maps from genus $g$ Riemann surfaces
with $h$ boundary components to ${\widetilde X}_\wm$ mapping the
boundary components to the exceptional curves $C_i$. The partition
function of the theory is a generating functional of the form
\be\label{eq:openA} \CF_{{\widetilde X}_\wm,\{C_i\},\{N_i\}} =
\sum_{g=0}^\infty \sum_{h_1,\ldots,h_v=0}^\infty g_s^{2g-2+h}
\CF_{{\widetilde X}_\wm,\{C_i\}}^{g,\{h_i\}} \prod_{i=1}^v N_i^{h_i}
\ee where $h=\sum_{i=1}^v h_i$. The coefficients $\CF_{{\widetilde
X}_\wm,C_i}^{g,\{h_i\}}$ are heuristically defined in terms of
functional integrals over maps from genus $g$ Riemann surfaces with
$h$ boundary components to ${\widetilde X}_m$ mapping $h_i$ boundary
components to the curve $C_i$. In the {\bf B}-model they are expected
to depend on complex moduli for the data $({\widetilde X}_\wm, C_i)$.
Since the exceptional curves are rigid in the threefold, it follows
that they only depend on the complex structure parameters of
${\widetilde X}_\wm$.

According to \cite{EW:CS}, the open-string path integral localizes on
degenerate maps, just as in the closed string situation. However,
degenerate open string maps collapse a Riemann surface with boundary
considered as a ribbon graph, to the corresponding graph embedded in
${\widetilde X}_\wm$.  In particular, the degenerate maps can have a nontrivial
structure even at genus zero. For this reason there is no rigorous
mathematical formulation of open-closed amplitudes except for $g=0$
and $h=0,1$ when degenerate maps are constant maps.  In this paper we
will focus only on these two cases.

The term corresponding to $g=h=0$ in \eqref{eq:openA} is the genus
zero closed string free energy $\CF^0_{{\wX}_{\wm}}$ which was
introduced in section \ref{ss:closed}. The term
corresponding to $g=0$, $h=1$ represents the disc open string free
energy which is determined by $3$ dimensional chain integrals of the
holomorphic three-form $\Omega_{\wX_{\wm}}$. To explain this
construction recall that we are considering D-brane configurations
satisfying the homology constraint \eqref{eq:gauss}. For a fixed
$\wm$, the solutions to this equation are in one-to-one correspondence
with  points in the lattice
\[
\Lambda_\wm =\hbox{ker}\left(H_2(C) \to H_2(\wX_\wm)\right)
\]
where $C=\cup_{i=1}^v C_i$.  Recall that in our set-up, the points in
$\bM$ parameterize threefolds with isolated ODP singularities, so that
the cover $\wbM\to \bM$ is unramified. Then the lattices $\Lambda_\wm$
span a locally constant sheaf when $\wm$ varies in $\wbM$. A flat section
of this sheaf parameterizes a pair $(\wX_\wm, C_{N,\wm})$ where
$C_{N,\wm}$ is a homologically trivial algebraic cycle on $\wX_\wm$ of
the form $C_{N,\wm}=\sum_{i=1}^v N_i C_i$.

Regarding $C_{N,\wm}$ as a topological brane on $\wX_\wm$, the disc
partition function is a period of the holomorphic three-form
$\Omega_{\wX_\wm}$ over a 3-chain, defined as follows.  Consider the
relative homology sequence for a pair $(\wX_\wm, C)$, with $\wX_\wm$
smooth
\[
%%\label{eq:homseqA}
0\to H_3(\wX_\wm, \IZ) \to
H_3(\wX_\wm,C,\IZ)\stackrel{\delta}{\ra} \Lambda_\wm \to 0.
\]
Given
a homologically trivial algebraic cycle $C_{N,\wm}$ on $\wX_\wm$, we
can find a relative cycle ${\widetilde \Gamma}_{N,\wm}\in
H_3(\wX_\wm,C)$ so that
\[
%%\label{eq:bound}
\delta {\widetilde
\Gamma}_{N,\wm} = C_{N,\wm}
\]
According to \cite{EW:branes,AV:discs}, for fixed $N_i$ satisfying the condition \eqref{eq:gauss}
the disc partition function is the
holomorphic function (= the Griffiths normal function associated to
$C_{N,\wm}$) on the enlarged moduli space $\wbM'$ given by
\be\label{eq:suppotA}
{\wCF}_{N,\wm} = \frac{1}{g_{s}} \sum_{i=1}^{v}
\mathcal{F}_{\wX_{\wm},\{ C_{i} \}}^{0, e_{i}}N_{i}
= \int_{\wGamma_{N,\wm}}
\Omega_{\wX_\wm}.
\ee
Here as usual we write $e_{i}$  for the $i$-th vector in the standard
basis of the lattice $\mathbb{Z}^{\oplus v}$.

Note that the relative cycle $\wGamma_N$ is
only determined modulo elements in $H_3(\wX_\wm, \mathbb{Z})$, therefore the
superpotential \eqref{eq:suppotA} is defined modulo periods.  In
general there is no preferred choice for a three-chain $\wGamma_N$,
hence we will regard this as a discrete ambiguity in the disc
partition function.

\subsection{Duality}

Large $N$ duality predicts a correspondence between topological open
and closed strings on a pair of Calabi-Yau threefolds related by an
extremal transition. A precise mathematical statement has not been
formulated for ${\bf B}$-model transitions between compact Calabi-Yau
threefolds.  On general grounds, large $N$ duality is expected to
relate genus zero open string amplitudes with $h$ boundary components
on $\wbM'$ to terms of order $h$ in the Taylor expansion of the closed
string partition function on $\bL'$ near $\bM'$. In this subsection
our goal is to formulate a precise mathematical statement for genus zero
terms with $h=0,1$.

We will show that for a certain choice of special coordinates
$z^\alpha$, $\alpha=1,\ldots,h^{1,2}(X_l)+1$ in a neighborhood of
$\bM'$ in $\bL'$, the closed string prepotential $\CF^0$ and its
derivatives $\left(\frac{\partial \CF^0}{\partial z^\alpha}\right)$
have well defined limits $\CF^0_{\bM'}$, $\left(\frac{{\partial
\CF^0}}{\partial z^\alpha}\right)_{\bM'}$ along $\bM'$. Moreover, the
following relations hold \be\label{eq:dualityrelations} {\widetilde
{\CF^0}} = {\rho'}^* (\CF^0_{\bM'}),\qquad \wCF_{i} = {\rho'}^*
\left(\frac{{\partial \CF^0}}{\partial z^i}\right)_{\bM'} \ee where
${\widetilde {\CF^0}}$ denotes the closed string prepotential on the
moduli space ${\widetilde {\bM'}}$, ${\widetilde \CF}^i$,
$i=1,\ldots,v-r$ are open string superpotentials of the form
\eqref{eq:suppotA}, and $i=1,\ldots,v-r$ labels the normal directions
to $\bM'$ in $\bL'$.  The main point we would like to make is that
these relations are a corollary of the following intrinsic geometric
results which will be proved in the next section.

\begin{description}
\item[(i)] For any $\wm\in \wbM$, and $m\in \bM$ so that
  $m=\rho(\wm)$, the contraction map \linebreak $(\wX_{\wm},C) \to
  (X_{m},\op{Sing}(X_{m}))$ induces  an
isomorphism
\[
H_3(\wX_\wm,C,\IZ) \simeq H_3(X_m,\IZ).
\]
Choose a retraction map
$\ret : \mycal{X} \to \mycal{X}_{|\bM}$ from the universal family on
$\bL$ onto the universal family on $\bM$ (This can always be done in a
neighborhood of a point $m \in \bM$.) Suppose  $\Gamma$ is a
multivalued section of the sheaf of
third homology groups $H_3(X_l,\mathbb{Z})$ on $\bL$, then the image of
$\Gamma$ under $\ret$
induces a multivalued section $\wGamma$ of the
sheaf of relative homology groups $H_3(\wX_\wm, C;\mathbb{Z})$ on $\wbM$.
\item[(ii)] Let
\[
\Pi = \int_{\Gamma_l} \Omega_{X_l},\qquad {\widetilde \Pi} =
\int_{\wGamma_{\wm}} \Omega_{\wX_\wm}
\]
be the periods of the respective holomorphic 3-forms, viewed as
multivalued holomorphic functions on $\bL'_0$ and $\wbM'$.
Then $\Pi$ induces a multivalued holomorphic function $\Pi_{\bM'}$  on
$\bM'$ and
\be\label{eq:limit}
{\widetilde \Pi} = {\rho'}^*(\Pi_{\bM'}).
\ee
\item[(iii)] For any points $l\in\bL_0$, $\wm\in \wbM$, such that
  $\ret(l) = \rho(\wm)$, there is a commutative diagram of the form
\be\label{eq:homdiagA} \xymatrix{ 0 \ar[r] &V_l^\perp /V_l
  \ar[d]^{\simeq} \ar[r] & H_3(X_l,\IZ)/V_l
  \ar[d]^{\simeq}\ar[r]^{\qquad p} & V_l ^{\vee} \ar[r]
  \ar[d]^{\simeq} & 0 \\ 0 \ar[r] & H_3(\wX_\wm,\IZ) \ar[r] &
  H_3(\wX_\wm,C,\IZ)\ar[r]^{\qquad \delta} & N_{\wm} \ar[r] & 0}
\ee
  where the rows are exact and the vertical arrows are isomorphisms.
  $V_l$ is the lattice of vanishing cycles in $H_3(X_l)$ and
  $V_l^\perp$ denotes the orthogonal lattice with respect to the
  intersection pairing. The map $p: H_3(X_l,\IZ)/V_l\to V_l^{\vee}$ is
  given by
$
p(\overline{\Gamma}) = \langle \overline{\Gamma}, \bullet \rangle
$
which is well defined because $V_l$ is isotropic with respect to the
  intersection pairing.
\end{description}

\

The claim {\bf (i)} follows from the definition of relative homology
groups. The proof of assertions {\bf (ii)} and {\bf (iii)} will be
given in section~\ref{s:mhs}.  In the remaining part of this section
we will explain how these statements lead to large $N$ duality up to
first order.  Recall from Section \ref{ss:closed} that for $l\in \bL_0$
the genus zero closed string partition function can be written locally
in terms of splitting of the third homology groups $H_3(X_l,\IZ)$ into
a direct sum of two Lagrangian sublattices.
Restricting to an open subset $\CU'$ of $\bL'$,
suppose we can choose the Lagrangian sublattices $A_l, B_l$ for
a smooth $X_l$ so that
\be\label{eq:dualityA} A_l = V_l \oplus
\widetilde{A}_l, \qquad B_l = U_l \oplus \widetilde{B}_l \ee
subject
to the following conditions:
\begin{itemize}
\item[a)] $A_l$ is contained in the fixed locus of the monodromy
  transformations \eqref{eq:monodromyA}.
\item[b)] $V_l$ is orthogonal to $\widetilde{B}_l$ and $U_l$ is
orthogonal to $\widetilde{A}_l$ with respect to the intersection
pairing.
\end{itemize}

\

\noindent
More concretely, this means that we choose the symplectic basis of
cycles so that \linebreak $\{\gamma_{1,l}, \ldots,
\gamma_{v-r,l}\}$ generate the lattice $V_l$ of vanishing cycles. Then
$U_l$ is generated by the dual cycles
$\{\gamma^1_l, \ldots, \gamma^{v-r}_l\}$ and  $V_l^\perp$ is generated by
$\{\gamma_{\beta, l}, \gamma^{\alpha}_l\}$ for $\alpha=v-r+1,\ldots,
h^{1,2}(X_l)+1$ and $\beta = 1, \ldots, h^{1,2}(X_l)+1$.
Although there is no such canonical choice, note that we can obtain a
basis of $V_l$ by choosing any collection $\bD_{i_1}, \ldots,
\bD_{i_{v-r}}$ of local components of the
discriminant which intersect transversely along $\bM$. Then we can
complete this basis to a symplectic basis
of cycles of $H_3(X_l,\IZ)$.

Given such a basis of cycles on each smooth $X_l$, the commutative
diagram \eqref{eq:homdiagA} shows that the cycles $\{\gamma_{\alpha,
l}, \gamma^{\alpha}_l\}$ for $\alpha=v-r+1,\ldots, h^{1,2}(X_l)+1$
induce a symplectic basis of cycles on $\wX_\wm$ for $\wm$ in the open
set ${\widetilde \CU}=\rho^{-1}(\CU\cap \bM)$. Moreover, to each basis
element $\gamma_{l}^{i}$, $i=1,\ldots, v-r$ we can associate an
exceptional curve $C_{i,\wm}$ on $\wX_\wm$ so that the $C_{i,\wm}$
generate $\Lambda_\wm$ for any $\wm\in {\widetilde \CU}$.  The images of the
cycles $\{\gamma^1_l, \ldots, \gamma^{v-r}_l\}$ in $H_3(V_l)/V_l$
induce relative three-cycles $\wGamma_{1,\wm}, \ldots,
{\wGamma}_{v-r,\wm}$ in $H_3(\wX_\wm, C)$ so that
\[
\delta (\wGamma_{i,\wm})= C_{i,\wm}
\]
for $i=1,\ldots, v-r$.
Then we have a well defined closed string prepotential
$\wCF:\widetilde{\CU}' \to \IC$
and open string superpotentials $\wCF_i : {\widetilde \CU}'\to \IC$
\[
\wCF_{i} = \int_{\wGamma_{i,\wm}}\Omega_{\wX_\wm}
\]
for each $C_{i,\wm}\in \Lambda_\wm$.  The special geometry relations
\eqref{eq:specialB}-\eqref{eq:specialC} together with statement {\bf
(ii)} above imply that the closed string prepotential $\CF$ and its
derivatives $\frac{\partial \CF}{\partial z^\alpha}$ have well defined
limits $\CF_{\bM'}$, $ \left(\frac{\partial \CF}{\partial
z^\alpha}\right)_{\bM'}$ at the points of  $\CU'\cap \bM'$.
The relations \eqref{eq:dualityrelations} follow from equation
\eqref{eq:limit}. Note that these relations are valid for
any choice of a symplectic basis in $H_3(X_l)$ satisfying conditions
$(a)$ and $(b)$ above.

\section{Geometric Transitions and Mixed Hodge Structures} \label{s:mhs}

In this section we would like to prove statements {\bf (ii)} and {\bf
(iii)}.  We will consider the following abstract situation. Let
$\pi:\CX\to \Delta$ be a one parameter family of projective Calabi-Yau
threefolds over the unit disc $\Delta$.  The generic fiber
$X_\mu$, $\mu \in \Delta \setminus\{0\}$ is assumed to be smooth and
the central fiber $X_0$ is a nodal threefold with ordinary double
points $p_1,\ldots, p_v$. We will assume that over the complement of the
central fiber, the family
$\pi:\CX\to \Delta$ is equipped with a globally defined relative
holomorphic 3-form, which restricts to a holomorphic volume form on
each fiber.  Moreover we will assume that there is a
smooth crepant projective resolution $\wX\to X_0$. Such a family can
be obtained for example by taking $\Delta$ to be a holomorphic disc in
the moduli space $\bL$ intersecting $\bM$ transversely at the origin.

The restriction of the family $\CX\to \Delta$ to the punctured disc
$\Delta^{\times}=\Delta \setminus\{0\}$ is a family of smooth Calabi-Yau
threefolds which determines a geometric variation of Hodge
structures.  To this data we can associate a period map $\phi :
\Delta^{\times} \to D/\mathfrak{M}$ \cite{PG:variations}.  where $D$ is the
classifying space of Hodge structures and $\mathfrak{M}$ is the
monodromy group of the family.

The proof of claim {\bf (ii)} is Hodge theoretic and is based on the
nilpotent orbit theorem and the Clemens-Schmid exact sequence
associated to the degeneration $\CX\to \Delta$. In the process of
proving claim {\bf (ii)}, we will also show that the Clemens-Schmid exact
sequence implies claim {\bf (iii)}.

First observe that the nilpotent orbit theorem implies the existence
of a well defined limit of the periods of the holomorphic three-form
at the origin $0\in \Delta$.  Suppose $\Gamma$ is a multivalued section
of the sheaf of third homology groups $H_3(X_s)$ over
$\Delta^{\times}$. Let
\[
%%\label{eq:periodA}
\Pi(s) = \int_{\Gamma_s}
\Omega_{X_s}
\]
be the period of the holomorphic three-form on
$\Gamma_s$, for $s\neq 0$. $\Pi$ is a multivalued holomorphic function
on the punctured disc $\Delta^{\times} = \Delta\setminus \{0\}$.  Let
\[
\psi: \Delta^* \to D^{\vee}, \quad \psi(s) =
\hbox{exp}\left(-\frac{{\log} s}{2\pi i}N\right) \phi(s),
\]
be the modified the period map $\phi$ with
values in the compact dual $D^{\vee}$ of $D$
\cite{PG:variations,WS:variation}.  Here $N= \log(T) = T - \op{id}$ is the
logarithm of the monodromy transformation about the origin. The
identity $\log(T) = T - \op{id}$ is equivalent to $(T - \op{id})^{2} = 0$
which holds for any ODP degeneration of threefolds. Indeed, this
follows from the observation that if we write $T-\op{id}$  as a block
matrix with respect to the decompositions
\eqref{eq:dualityA}, then the only non-zero block of $T-\op{id}$ is
the one sending $U_{l}$ to $V_{l}$.

According
to the nilpotent orbit theorem \cite{WS:variation}, the map $\psi$ can
be extended to a
single valued holomorphic map
\[
\overline{\psi} : \Delta \to D^{\vee}
\]
In particular, the multivalued map
$\Pi:\Delta^{\times} \to \IC$, which is a matrix coefficient of
$\phi$,  can be written as
\[
\Pi(s) = {\overline \Pi}(s) + \frac{1}{2\pi i} (\hbox{log s}) \eta(s)
\]
for some $\overline{\Pi}(s)$, $\eta(s)$ single valued holomorphic
functions of $s$ with $\eta(0)=0$.  Later in the proof of relation
\eqref{eq:limit} we will see that the value $\overline{\Pi}(0)$ admits
an intrinsic description of in terms of the mixed Hodge structure of
the semistable model of the degeneration $\CX\to \Delta$.

The point $\overline{\psi}(0)\in D^{\vee}$ corresponds to the limiting
mixed Hodge structure on the third cohomology group of a smooth fiber
$H^3(X_s)$, $s\neq 0$. On the other hand the relative cohomology group
$H^3(\wX,C)$ also carries a canonical mixed Hodge structure
\cite{D:HodgeII,D:HodgeIII}. The relation
\eqref{eq:limit} that we wish to prove asserts that a period in the
limiting mixed Hodge structure on $H^3(X_s)$ is equal to the corresponding
period in the mixed Hodge structure on  $H^3(\wX,C)$. In particular
\eqref{eq:limit} will follow if we can show that the natural map
$H^{3}(\wX,C) \to H^{3}(X_{s})$ is an inclusion of mixed Hodge
structures.

We claim that this follows from the Clemens-Schmid exact sequence.  In
order to formulate a more precise statement, let us first describe the
mixed Hodge structure on $H^3(\wX,C)$ and construct the Clemens-Schmid
exact sequence for our degeneration.

Recall that a mixed Hodge structure on a
vector space $H=H_\IZ \otimes \IC$ is defined by

\begin{itemize}
\item a descending Hodge filtration $\{F^k\}$, and
\item an ascending weight filtration $\{W_m\}$, defined over
  $\mathbb{Q}$,
so that for every $m$ the successive quotient $W_m/W_{m-1}$ has a
  pure Hodge structure of weight $m$ given by the induced
filtration
\[
F^k(W_m/W_{m-1}) = (W_m\cap F^k)/(W_{m-1}\cap F^k).
\]
\end{itemize}

\

\noindent
The limiting mixed Hodge structure on the cohomology of a smooth fiber
is defined by the Hodge filtration given by the nilpotent orbit
\cite{WS:variation} and by the monodromy weight filtration
\[
0\subset W_0\subset W_1\subset \cdots \subset W_{5} \subset W_{6}=
H^3(X_s).
\]
which is determined by the monodromy action on cohomology.
In our case, $N^2=0$, and the monodromy weight filtration is very simple
\[
%%\label{eq:monfiltr}
W_3 = \hbox{ker}(N), \qquad W_2=\hbox{im}(N).
\]
 Additional
information about the limiting mixed Hodge structure can be obtained
from the Clemens-Schmid exact sequence as we will see below.

To describe the mixed Hodge structure on $H^3(\wX,C)$, consider the
long exact sequence of the pair $(\wX,C)$
\[
%%\label{eq:longseq}
\cdots \to H^2(\wX) \to H^2(C) \to H^3(\wX,C)\to H^3(\wX) \to 0
\]
where all cohomology groups have complex coefficients.
The ascending weight filtration on $H^3(\wX,C)$ is defined  by
\[
%%\label{eq:weight}
\begin{aligned}
& W_3 = H^3(\wX,C)\cr
& W_2 = \hbox{Im}\left(H^2(C) \to H^3(\wX,C)\right)\simeq
  H^2(C)/\hbox{im}\left(H^2(\wX)\to H^2(C)\right)\cr
& W_1=0.\cr
\end{aligned}
\]
The Hodge filtration is the standard Hodge filtration on the cohomology of the quasi-projective variety
$X\setminus C$.
The relation between the mixed Hodge structures on $H^{3}(X_{s})$ and
$H^3(\wX,C)$ will be extracted from  the
Clemens-Schmid exact
sequence, which is constructed in terms of a semistable model of the
degeneration
$\CX\to \Delta$.

A semistable model for the family $\CX\to \Delta$ is a new family
$\oCX\to \Delta$ which fits in a commutative diagram of the form
\be\label{eq:semistableA} \xymatrix{ \oCX \ar@{-->}[r]\ar[dr] & f^*\CX
\ar[d] \ar[r] & \CX\ar[d] \\ & \oDelta \ar[r]^f & \Delta } \ee Here
$\overline{\mathcal{X}}$ is assumed smooth, $f:\oDelta \to \Delta$ is
a finite cover of $\Delta$ branched at the origin and the dashed arrow
represents a birational map which is an isomorphism over the punctured
disc $\oDelta^{\times}$. In addition, the central fiber $\oX\equiv
\oX_0$ is required to be a normal crossing divisor in
$\overline{\mathcal{X}}$ with smooth reduced irreducible components.

The cohomology of the central fiber $\oX_0$ of the semistable
degeneration can be equipped with a mixed Hodge structure, which can
be described in terms of the Mayer-Vietoris spectral sequence.  The
Clemens-Schmid exact sequence is an exact sequence of mixed Hodge
structures of the form
\[
%%\label{eq:CSa}
\cdots \to H_1(\oX_{0}) \to
H^3(\oX_{0}) \to H^3(\oX_\os) \stackrel{N}{\ra} H^3(\oX_\os) \to
H_3(\oX_{0}) \to \cdots
\]
The Clemens-Schmid theorem \cite{C:degeneration} guarantees
that all maps in this sequence are
morphisms of mixed Hodge structures. The mixed Hodge structure on
homology is induced by the mixed Hodge structure on cohomology using
the universal coefficient formula.

Now, claim {\bf (ii)} follows from the following two facts which will
be proven in the rest of the section.
\begin{itemize}
\item $H^3(\oX_0)\simeq \hbox{ker}\left(N:
 H^3(\oX_\os) \to H^3(\oX_\os)\right)$ as mixed Hodge structures. This
 follows from $H_1(\oX_{0})=0$.
\item  $H^3(\wX,C) \simeq H^3(\oX_{0})$ as mixed Hodge structures.
\end{itemize}
To begin with, let us construct the semistable degeneration for our
family $\CX\to \Delta$.  Let $f:{\overline \Delta} \to \Delta$ be a
double cover of the disc defined in local coordinates by $s={\overline
s}^2$, and let $f^*\CX$ denote the pull-back of the family $\CX\to
\Delta$ to ${\overline \Delta}$. The total space of the family
$f^*\CX\to {\overline \Delta}$ has double point singularities at the
points $p_i$ on the central fiber $(f^*\CX)_0$, which is canonically
isomorphic to $X_0$. We construct a new family ${\overline \CX}\to
{\overline \Delta}$ by blowing-up the singular points on the total
space of $f^*\CX$. Since all the singular points lie on the central
fiber, it is an easy check that the fiber ${\overline X}_{\overline s}$
is isomorphic to $X_s$, $s={\overline s}^2$, for ${\overline s}\neq
0$. The central fiber $\overline{X}_0$ is a normal crossing
variety consisting of $v+1$ smooth reduced irreducible components
\[
\overline{X}_{0} = \widehat{X}\cup Q_1\ldots\cup Q_v
\]
where $\widehat{X} \to X_0$ is the blow-up of $X_0$ at the $v$
singular points and $Q_1, \ldots, Q_v$ are quadric threefolds. The
blow-up $\widehat{X}\to X_0$ replaces each ordinary double point $p_i$
with an exceptional divisor $E_i$ isomorphic to the Hirzebruch surface
$\IF_0$. $\widehat{X}$ intersects each quadric threefold $Q_i$ along
$E_i$, which is a hyperplane section of $Q_i$. Note that the
exceptional divisors $E_i$ are pairwise disjoint, hence the components
$Q_i$ have no common points. Therefore $\overline{X}_{0}$ is indeed a
normal crossing divisor with smooth reduced irreducible components.

Note that a small resolution of $X_0$ can be obtained by contracting
${\widehat X}$ along a collection of rulings of the exceptional
divisor $E_i\subset {\widehat X}$.  Since each divisor $E_i$ admits
two distinct rulings, we can obtain in principle different small
resolutions $\wX$ related by flops. Note however that not all possible
contractions result in projective small resolutions. The
considerations of this section are valid for any projective
contraction ${\widehat X}\to \wX$.

Next, we will compute the rational homology of the semi-stable model
$\oX_{0}$ using Mayer-Vietoris exact sequences. First we compute the
homology of the singular fiber $X_0$ in the initial family.  From a
topological point of view $X_0$ can be represented as the cone of a
map from a disjoint collection of $v$ three-spheres to a smooth fiber
$X_s$
\[
f:\sqcup_{i=1}^v S^3_i \to X_s
\]
so that $f$ maps each $S^3_i$ homeomorphically onto a vanishing cycle
associated to the $i$-th node of $X_0$.  Therefore $X_0$ is homotopy
equivalent to the union of $X_s$ and $v$ closed four-discs $D^4_i$,
$i=1,\ldots, 4$ so that $X_s\cap D^4_i= f(S^3_i)$. The associated
homology Mayer-Vietoris sequence reads
\be\label{eq:MVa}
\cdots \to
\oplus_{i=1}^v H_k(S^3_i)\to H_k(X_s) \oplus \oplus_{i=1}^v H_k(D^4)
\to H_k(X_0) \to \oplus_{i=1}^v H_{k-1}(S^3_i)\to\cdots
\ee
In
particular we have
\[
%%\label{eq:MVb}
0\to H_4(X_s) \to H_4(X_0) \to
\oplus_{i=1}^v H_{3}(S^3_i)\to H_3(X_s) \to H_3(X_0) \to 0
\]
The image of the map $\oplus_{i=1}^v H_{3}(S^3_i)\to H_3(X_s)$ is the
$v-r$ dimensional
lattice $V$ of vanishing cycles on $X_s$, and the kernel is the
$r$-dimensional lattice $R$ of relations among vanishing
cycles. Therefore we find two short
exact sequences
\[
%%\label{MVc}
\begin{aligned}
& 0\to H_4(X_s) \to H_4(X_0) \to R \to 0 \cr
& 0 \to V \to H_3(X_s) \to H_3(X_0) \to 0 \cr
\end{aligned}
\]
which yield the relations $b_3(X_0) = b_3(X_s)-(v-r)$,
$b_4(X_0)=b_4(X_s) + r$.
The remaining Betti numbers can be easily determined by writing down
the other terms in \eqref{eq:MVa}.
We record the complete results below, using the notation $b_k\equiv
b_k(X_s)$
\be\label{eq:bettiA}
\begin{tabular}{||l||c|c|c|c|c|c|c||} \hline
$i$ & $6$ & $5$ & $4$ & $3$ & $2$ & $1$ & $0$ \\ \hline\hline
$b_{i}(X_{s})$ & $1$ & $0$ & $b_{2}$ & $b_{3}$ & $b_{2}$ & $0$ & $1$ \\ \hline
$b_{i}(X_{0})$ & $1$ & $0$ & $b_{2}+r$ & $b_{3} - (v - r)$ & $b_{2}$ &
  $0$ & $1$
 \\ \hline
\end{tabular}
\ee
In order to determine the homology of the blow-up ${\widehat X}$,
consider the long exact homology sequence
for the pair $({\widehat X}, E)$
\be\label{eq:blowupA}
\cdots \to H_k(E)\to H_k({\widehat X}) \to H_k({\widehat X}, E)\to
H_{k-1}(E)\to \cdots,
\ee
where $E=\cup_{i=1}^v E_i$ denotes the
exceptional divisor.
Using the fact that $E_i \simeq \IP^1\times \IP^1$ for each
$i=1,\ldots, v$ and that $X_{0}$ is $\widehat{X}$ with each $E_{i}$
collapsed to a point, we see that
\eqref{eq:blowupA} reduces to
the following straightforward isomorphisms
\[
%%\label{eq:blowupB}
\begin{aligned}
H_k({\widehat X}) \simeq H_k(X_0), \ k=0,1,5,6
\end{aligned}
\]
and two exact sequences of the form
\be\label{eq:blowupC}
\begin{aligned}
& 0\to H_4(E) \to H_4({\widehat X}) \to H_4({\widehat X}, E)\to 0\cr
& 0\to H_3({\widehat X})\to H_3({\widehat X}, E)\stackrel{\delta}{\ra}
H_2(E)\stackrel{\iota_*}{\ra} H_2({\widehat X}) \to H_2({\widehat X}, E)
\to 0.\cr
\end{aligned}
\ee
The first sequence yields
\[
b_4({\widehat X}) = b_4(X_0) + v = b_2(X_{s}) + v + r.
\]
By Poincar\'e duality we also have
\[
b_2({\widehat X})=b_4({\widehat X}) =b_2(X_{s}) + v + r.
\]
Also taking into account \eqref{eq:bettiA} and the second sequence in
\eqref{eq:blowupC} we find
\[
%%\label{eq:blowupD}
b_3({\widehat X}) =b_3(X_s) -2(v-r).
\]
Moreover, it follows that
\[
%%\label{eq:blowupE}
\hbox{dim}(\hbox{ker}(\iota_*))=v-r,\quad \hbox{dim}(\hbox{im}(\iota_*))= v+r
\]
where $\iota_*: H_2(E)\to H_2({\widehat X})$ is the map
on homology determined by the
inclusion $\iota:E\to {\widehat X}$.
Note that $\hbox{ker}(\iota_*)=\hbox{im}(\delta)$ is spanned by
relative homology three-cycles for
the pair $({\widehat X}, E)$ modulo cycles in ${\widehat X}$. These
cycles generate the lattice
of relations among
curve classes on $E$ regarded as homology cycles on ${\widehat X}$.

There is a similar exact sequence for the pair $(\wX,C)$, where $\wX$
is a small projective
resolution of $X_0$ and $C$ is the collection of exceptional curves
\be\label{eq:smallA}
\cdots \to H_k(C)\to H_k({\wX}) \to H_k({\wX}, C)\to H_{k-1}(C)\to \cdots
\ee
Again, by construction we have $C_{i} \cong \mathbb{P}^{1}$ and
$X_{0}$ is $\wX$ with each $C_{i}$ collapsed to a point.
The long exact sequence \eqref{eq:smallA} reduces to isomorphisms
\be\label{eq:smallB}
H_k(\wX) = H_k(X_0),\ k=0,1,4,5,6,
\ee
and a shorter exact sequence
\be\label{eq:smallC}
\begin{aligned}
0\to H_3(\wX) \to H_3(\wX,C) \stackrel{\widetilde \delta}{\ra}
H_2(C)
\stackrel{{\widetilde \iota}_*}{\ra} H_2(\wX) \to H_2(\wX,C) \to
0.
\end{aligned}
\ee
Then, using \eqref{eq:smallB}, \eqref{eq:smallC} and the fact that
$b_{2}(\wX) = b_{4}(\wX)$ one gets
\[
%%\label{eq:smallD}
b_3(\wX) = b_3(X_s)-2(v-r).
\]
and
\[
%%\label{eq:smallE}
\hbox{dim}(\hbox{ker}({\widetilde \iota}_*))=v-r,\quad
\hbox{dim}(\hbox{im}({\widetilde \iota}_*))= r
\]
where ${\widetilde \iota}_*: H_2(C) \to H_2(\wX)$ is the
map on homology determined
by the inclusion ${\widetilde \iota}: C\to \wX$. Note that
$\hbox{ker}({\widetilde \iota}_*)\simeq \hbox{im}({\widetilde
  \delta})$ is spanned by relative
homology three-cycles on $(\wX,C)$ modulo cycles in $\wX$. Just as in
the blow-up case
these cycles generate the lattice of relations among exceptional curve
classes on $\wX$.

For future reference, we analyze the direct relation between the
rational homology of ${\widehat X}$ and the homology of $\wX$. Choose
a basis of homology two-cycles $(a_i, b_i)$ in
$H_2(E_i)$, $i=1,\ldots, v$ corresponding to the two rulings so that
the contraction map $q:{\widehat X}\to \wX$ contracts all rational
curves in the classes $b_i$, $i=1,\ldots, v$. A simple intersection
computation yields the following relations in the homology ring of
${\widehat X}$ \be\label{eq:inters} a_i\cdot E_j = b_i \cdot E_j =
-\delta_{ij}, \quad i,j =1,\ldots, v.  \ee Now let us consider the
following commutative diagram of spaces
\[
%%\label{eq:commdiagA}
\xymatrix{ E \ar[r]^{\iota} \ar[d]_p & {\widehat X}\ar[d]^q\\
  C\ar[r]^{\widetilde \iota} & \wX\\}
\]
where $p:E\to C$ is a projection map
contracting the rulings $b_i$, $i=1,\ldots, v$.  Then we have a
natural relation between maps on homology
\[
q_*\iota_* = {\widetilde \iota}_*p_*.
\]
This shows that that $p_*:H_2(E)\to H_2(C)$ induces a map
\[
%%\label{eq:kerisom}
{\overline p}_*: \hbox{ker}({\iota}_*) \to \hbox{ker}({\widetilde \iota}_*).
\]
Moreover, one can show that ${\overline p}_*$ is injective, since any
element in the kernel
of ${p}_*$ is necessarily a linear combination of $b_i\in
H_2(E)$. However, any nontrivial
linear combination of $b_i$ cannot lie in $\hbox{ker}(\iota_*)$
because the intersection pairing on
$(b_i, E_i)$, $i=1,\ldots, v$ is nondegenerate according to
\eqref{eq:inters}. Therefore
$\hbox{ker}({\overline p}_{*})=0$. Since
$\hbox{ker}({\iota}_{*}),\hbox{ker}({\widetilde \iota}_{*})$ are
${\mathbb Q}$-vector spaces of equal dimension $v-r$, it follows that
${\overline p}_*$ is an isomorphism. Furthermore,
$\hbox{ker}(\iota_*)$ is contained in the linear span
of the homology classes $a_i-b_i$, $i=1,\ldots, v$.

Note also that the map $q_*:H_3(\hX, E)\to H_3(\wX, C)$ is an
isomorphism. This follows from the
commutative diagram of pairs
\[
\xymatrix{ (\hX,E) \ar[d]_q \ar[dr] \cr (\wX,C) \ar[r] & (X_0,
  X_0^{\text{sing}})\\}
\]
which induces isomorphisms $H_3(\hX,E) \simeq H_3(X_0,X_0^{sing})$ and
$H_3(\wX,C)\simeq
H_3(X_0,X_0^{sing})$. Therefore we obtain the following commutative
diagram of homology groups
\be\label{eq:commdiagB}
\xymatrix{
0 \ar[r]& H_3(\hX) \ar[r]\ar[d]_{q_*} & H_3(\hX,E) \ar[r]^\delta
\ar[d]_{q_*} &  \hbox{ker}(\iota_*)
\ar[r]\ar[d]_{{\overline p}_*} & 0 \\
0 \ar[r]& H_3(\wX) \ar[r]& H_3(\wX,C) \ar[r]^{\widetilde \delta} &
\hbox{ker}({\widetilde \iota}_*)
\ar[r]& 0\\}
\ee
where the rows are exact and the middle and rightmost vertical arrows
are isomorphisms. We conclude that the
leftmost vertical arrow must be an isomorphism as well, and there is
also an isomorphism
\[
%%\label{eq:relations}
H_3(\hX,E)/ H_3(\hX) \simeq H_3(\wX,C)/ H_3(\wX)
\]
To conclude the homology computations, let us determine the homology
of the central fiber $\oX_{0}$ of the
semistable degeneration. We will employ again a Mayer-Vietoris exact
sequence with respect to the
closed cover
\[
\oX_{0} = \hX \cup Q_1\cup,  \ldots, \cup Q_v
\]
where $Q_i$ are smooth quadric threefolds intersecting $\hX$
transversely along the exceptional
divisors $E_i$, $i=1,\ldots, v$. Let $Q$ denote the disjoint union
$Q=Q_1\cup \ldots \cup Q_v$.
Then, we have
\be\label{eq:MVd}
\begin{aligned}
0 & \to H_6(\hX)\oplus H_6(Q) \to H_6(\oX_{0}) \to H_5(E) \to
H_5(\hX)\oplus H_5(Q) \to H_5(\oX) \cr
& \to H_4(E) \to H_4(\hX)\oplus H_4(Q) \to H_4(\oX_{0}) \to H_3(E)\to
H_3(\hX)\oplus H_3(Q) \cr
& \to H_3(\oX) \to H_2(E) \to H_2(\hX)\oplus H_2(Q) \to H_2(\oX_{0})
\to H_1(E) \cr
& \to H_1(\hX)\oplus H_1(Q)
\to H_1(\oX) \to H_0(E) \to H_0(\hX)\oplus H_0(Q)\to H_0(\oX_{0})\to
0\cr
\end{aligned}
\ee
The homology of $Q$ can be computed easily from the Lefschetz
hyperplane theorem and an Euler characteristic computation via
Gauss-Bonet formula. We have
\[
H_k(Q)={\mathbb Q},\ k=0,2,4,6, \quad H_k(Q)=0, \ k=1,3,5.
\]
Moreover, since $E_i$ is a hyperplane section of $Q_i$, $i=1,\ldots,
v$,  it follows that the map
\[
H_4(E) \to H_4(\hX)\oplus H_4(Q)
\]
is injective. Thus the long exact
sequence \eqref{eq:MVd} yields the following straightforward
isomorphisms
\be\label{eq:sstableA}
\begin{aligned}
& H_6(\oX_{0}) \simeq H_6(\hX) \oplus H_6(Q) = {\mathbb Q}^{\oplus
    (1+v)}\cr
& H_5(\oX_{0}) \simeq H_5(\hX)\oplus H_5(Q) =0 \cr
& H_0(\oX_{0}) \simeq H_0(\hX) ={\mathbb Q} \cr
& H_1(\oX_{0}) \simeq H_1(\hX) =0\cr
\end{aligned}
\ee
The remaining part of \eqref{eq:MVd} splits into two exact sequences
\be\label{eq:sstableB}
\begin{aligned}
& 0 \to H_4(E) \to H_4(\hX)\oplus H_4(Q) \to H_4(\oX_{0}) \to 0 \cr
& 0 \to H_3(\hX) \to H_3(\oX_{0}) \to H_2(E) {\buildrel (\iota_*,j_*)\over \ra}
H_2(\hX)\oplus H_2(Q) \to H_2(\oX_{0}) \to 0 \cr
\end{aligned}
\ee
where the $j_*:H_2(E)\to H_2(Q)$ is the map on homology
determined  by
the inclusion $j:E\hookrightarrow Q$.
The first sequence gives
\[
b_4(\oX) = b_4(\hX) + v - v = b_{4}(\hX) = b_{2}(X_{s}) + v + r.
\]
Since $E$ is a hyperplane section of $Q$, the kernel of the map
$j_*:H_2(E)\to H_2(Q)$ is spanned by the homology classes $a_i-b_i$,
$i=1,\ldots, v$. On the other hand $\hbox{ker}(\iota_*)$ is a $v-r$
dimensional subspace of the linear span of $a_i-b_i$, $i=1,\ldots,
v$. Therefore we find that
\[
\hbox{ker} (\iota_*,j_*)\simeq \hbox{ker}(\iota_*)
\]
is also $v-r$ dimensional. This fixes the remaining Betti numbers
\[
%%\label{eq:sstableC}
\begin{aligned}
b_3(\oX) & = b_3(\hX)+(v-r)=b_3(X_s)-(v-r) \cr
b_2(\oX) & = b_2(\hX) +v -(2v-v+r) = b_2(X_s)+v.\cr
\end{aligned}
\]
In summary we can extend the table \eqref{eq:bettiA} to include all
the spaces
appearing in the geometric transition and the semi-stable
degeneration:
\begin{equation}
\label{eq:bettiall}
\begin{tabular}{||l||c|c|c|c|c|c|c||} \hline
$i$ & $6$ & $5$ & $4$ & $3$ & $2$ & $1$ & $0$ \\ \hline\hline
$b_{i}(X_{s})$ & $1$ & $0$ & $b_{2}$ & $b_{3}$ & $b_{2}$ & $0$ & $1$ \\ \hline
$b_{i}(X_{0})$ & $1$ & $0$ & $b_{2}+r$ & $b_{3} -(v - r)$ & $b_{2}$ &
  $0$ & $1$
 \\ \hline
$b_{i}(\hX)$ & $1$ & $0$ & $b_{2}+v+r$ & $b_{3} -2(v - r)$ & $b_{2}+v+r$ &
  $0$ & $1$
 \\ \hline
$b_{i}(\wX)$ & $1$ & $0$ & $b_{2}+r$ & $b_{3} -2(v - r)$ & $b_{2}+r$ &
  $0$ & $1$
 \\ \hline
$b_{i}(\oX_{0})$ & $1+v$ & $0$ & $b_{2}+v+r$ & $b_{3} -(v - r)$ & $b_{2}+v$ &
  $0$ & $1$
 \\ \hline
$b_{i}(\wX,C)$ & $1$ & $0$ & $b_{2}+r$ & $b_{3} -(v - r)$ & $b_{2}$ &
  $v-1$ & $1$
 \\ \hline
\end{tabular}
\end{equation}
With all this information in place, we are ready to finish the proof
of {\bf (ii)}.

Recall that
\[
\hbox{ker}(\iota_*)\simeq H_3(\hX,E)/H_3(\hX)
\]
according to the top exact sequence in \eqref{eq:commdiagB}. Then the
second exact sequence in
\eqref{eq:sstableB} yields a short exact sequence of the form
\begin{equation} \label{eq:barX0}
0\to H_3(\hX) \to H_3(\oX_{0}) \to  H_3(\hX,E)/H_3(\hX) \to 0.
\end{equation}
Comparing \eqref{eq:barX0} with the top row of \eqref{eq:commdiagB} we
get an isomorphism $H_{3}(\oX_{0}) \cong H_3(\hX,E)/H_3(\hX)$ of mixed
Hodge structures. Combined with the fact that the vertical arrows in
\eqref{eq:commdiagB} are isomorphisms we get an isomorphism
$H_{3}(\oX_{0}) \cong H_{3}(\wX,C)$ as mixed Hodge
structures. Dualizing and applying the universal coefficients theorem
we get the desired isomorphism of mixed Hodge structures
$H^{3}(\oX_{0}) \cong H^{3}(\wX,C)$. Since by \eqref{eq:sstableA} we
have $H_{1}(\oX_{0}) = 0$ this completes the proof of {\bf (ii)}.

At this point we still have to tie a few loose ends and prove claim
{\bf (iii)}. Note that the universal coefficients theorem and
\eqref{eq:bettiall} give that
$H^5(\oX_{0})=0$. Therefore the Clemens-Schmid exact sequence reduces to
\[
0 \to H^3(\oX_\os)/\hbox{im}(N) {\buildrel {\overline \beta} \over
  \ra} H_3(\oX_{0}) \to 0.
\]
The middle map $\overline{\beta}$
in this exact sequence is defined by the sequence of maps
\[
H^3(\oX_\os){\buildrel PD\over \ra} H_3(\oX_\os) \to H_3(\oCX) \simeq
H_3(\oX_{0})
\]
where the first map is Poicar\'e duality on a smooth fiber and the
next map is induced by inclusion in the
total space $\oCX$ of the semistable family. Since the image of
$N=\hbox{log}(T) = T-\text{id}$ is
the space $V$ of vanishing cycles, it follows that we have an
isomorphism of mixed Hodge
structures
\be\label{eq:mixedhom}
0\to H_3(\oX_\os)/V \to H_3(\oX_0) \to 0.
\ee
The weight filtration on $H_3(\oX_\os)/V$ is a one step filtration
induced by the monodromy weight
filtration of $H_3(\oX_\os)$
\[
\begin{aligned}
W_3 & = H_3(\oX_\os)/V\cr
W_2 & = (\ker N) /V = V^\perp/V\cr
W_1 & =0\cr
\end{aligned}
\]
Therefore we have a three-term exact sequence
\[
0 \to V^\perp/V \to  H_3(\oX_\os)/V {\buildrel p\over \ra} V^{\vee} \to 0
\]
where the map $p$ is induced by $N=\hbox{log}(T)= T-\op{id}$
\[
p({\overline \Gamma}) =\langle {\overline \Gamma}, \bullet \rangle.
\]
The monodromy weight filtration on $H_3(\oX_{0})$ is also a one step
filtration of the form
\[
\begin{aligned}
W_3 & = H_3(\oX_{0}) \simeq H_3(\wX,C)\cr
W_2 & = H_3({\widehat X}) \simeq H_3(\wX) \cr
W_1 & =0.\cr
\end{aligned}
\]
Since \eqref{eq:mixedhom} is an isomorphism of mixed Hodge structures,
it is in particular compatible with the weight filtrations. This
implies the commutative diagram \eqref{eq:homdiagA} of claim {\bf (iii)}.

%%\bibliographystyle{my-h-elsevier}
%%\bibliography{mhs}

\end{document}